\documentclass[preprint]{elsarticle}
\usepackage{graphicx}
\usepackage{amsmath,amssymb,amsfonts}

\begin{document}
\begin{frontmatter}

\title{The percolation phase transition and statistical multifragmentation in finite systems}

\author[1]{T.~Pietrzak}
\author[2,3]{A.~S.~Botvina}   
\author[1]{J.~Brzychczyk\corref{cor1}}
  \ead{janusz.brzychczyk@uj.edu.pl}
\author[4]{N.~Buyukcizmeci}  
\author[5]{A.~Le~F\`{e}vre}
\author[6]{J.~{\L}ukasik}
\author[6]{P.~Paw{\l}owski}
\author[7]{C.~Sfienti}
\author[5]{W.~Trautmann}  
\author[1]{A.~Wieloch}
\cortext[cor1]{Corresponding author}

\address[1] {Smoluchowski Institute of Physics, Jagiellonian University, Pl-30348 Krak\'ow, Poland}
\address[2] {Institute for Theoretical Physics, J. W. Goethe University, D-60438 Frankfurt am Main, Germany}
\address[3] {Institute for Nuclear Research, Russian Academy of Sciences, 117312 Moscow, Russia}
\address[4] {Department of Physics, University of Sel\c{c}uk, 42079 Konya, Turkey}
\address[5] {GSI Helmholtzzentrum f\"{u}r Schwerionenforschung GmbH, D-64291 Darmstadt, Germany}
\address[6] {H. Niewodnicza{\'n}ski Institute of Nuclear Physics, Pl-31342 Krak{\'o}w, Poland}
\address[7] {Institute of Nuclear Physics, Johannes Gutenberg University, D-55099 Mainz, Germany}

\begin{abstract}
The cumulant ratios up to fourth order of the $Z$ distributions of the largest fragment in spectator fragmentation 
following $^{107,124}$Sn+Sn and $^{124}$La+Sn collisions at 600 MeV/nucleon have been investigated. 
They are found to exhibit the signatures of a second-order phase transition established with cubic bond percolation and previously observed in the ALADIN experimental data for fragmentation of $^{197}$Au projectiles at similar energies. The deduced pseudocritical points are found to be only weakly dependent on the $A/Z$ ratio of the fragmenting spectator source. 
The same holds for the corresponding chemical freeze-out temperatures of close to 6 MeV.

The experimental cumulant distributions are quantitatively reproduced with the 
Statistical Multifragmentation Model and parameters used to describe the experimental fragment multiplicities, isotope distributions and their correlations with impact-parameter related observables in these reactions. 
The characteristic coincidence of the zero transition of 
the skewness with the minimum of the kurtosis excess appears to be a generic property 
of statistical models and is found to coincide with the maximum of the heat 
capacity in the canonical thermodynamic fragmentation model. 

\end{abstract}

\end{frontmatter}  


It has been shown in a recent paper that the statistical measures skewness and kurtosis excess of higher-order fluctuations of the largest fragment size
provide a robust indication of the transition point, linked to a second-order phase transition in the continuous limit~\cite{jb06}. 
The characteristic signatures identified with bond percolation calculations were found to be present in the ALADIN fragmentation 
data reported by Sch\"{u}ttauf {\it et al.}~\cite{sch96} for $^{197}$Au projectiles at 600 to 1000 MeV/nucleon in collisions with several targets~\cite{jb18}. 
Comparisons made with calculations for estimated system sizes demonstrated the high accuracy of the percolation model in 
reproducing the distributions of the largest fragment charge (atomic number $Z$) as well as the whole fragmentation 
pattern as represented by various forms of charge correlations. In analogy to percolation, the critical and pseudocritical 
points were identified in the fragmentation data~\cite{jb18}.

Very satisfactory reproductions of measured fragment yields and correlations had already been achieved in earlier studies of various kinds of proton and heavy-ion induced 
reactions~\cite{kreutz93,bauerbotv95,kleine2002}. In fact, the usefulness of percolation models for the interpretation of multi-fragment decay processes was realized soon after first experimental evidence (see, e.g., Refs.~\cite{minich82,jakob82,hirsch84})
for this new type of reaction mode became
available~\cite{bauer85,bauer86,campi86,biro86,campi88}. The results suggested that nuclei break up similar to percolation clusters, seemingly contradicting the first-order nature of the liquid-gas phase transition whose signatures were expected to appear in the new fragmentation data. As will be shown in the following, percolation exhibits phenomena that are generic for disintegration processes in finite systems, rather than permitting identifications of the transition order or the universality class.

The largest fragments and their charge distributions have received particular attention in nuclear multifragmentation studies. The largest fragment is expected to represent the liquid part of the system and thus to assume the role of the order parameter in a manifestation of the liquid-gas phase transition in finite systems, providing valuable insight into their phase behavior (for reports and reviews see, e.g., Refs.~\cite{jb18,gupta98,botet01,carmona02,frankland05,ma05,gulm2005,leneindre2007,das18,borderie19} and references given therein). 

In this work, we show that the observed properties of the distributions of the largest fragment charge are not restricted to the percolation transition but also observed at the liquid-vapor transition described with statistical multifragmentation models. The analysis will be applied to the ALADIN S254 experimental data obtained with stable and radioactive beams to investigate the isotopic
dependence of projectile fragmentation. The study was conducted with neutron-rich $^{124}$Sn and radioactive neutron-poor $^{107}$Sn and $^{124}$La beams of 600 MeV/nucleon incident energy and natural Sn targets~\cite{sfienti09,ogul11}. 

For the interpretation of the measured fragmentation processes and the observed isotopic effects, the Statistical 
Multifragmentation Model (SMM, Ref.~\cite{smm}) was applied~\cite{ogul11}. An ensemble of excited sources, defined with a 
few parameters, was used to represent the intermediate stage of the reaction at which equilibrium is assumed, an established method
successfully applied in previous interpretations of ALADIN fragmentation data~\cite{barz93,botvina95}.
The disintegration and further deexcitation of excited fragments is then followed within the statistical framework of the model. The experimental data were very well reproduced and, in particular, also the mean value of the largest fragment charge and its evolution with impact parameter were obtained with high accuracy. Here we present the cumulant ratios of the distributions of the largest fragment charge in order to investigate whether the experimental data and the SMM results equally exhibit the higher-order features observed with percolation. The SMM has been conceived to model the signatures of the nuclear liquid-gas phase transition, a first-order transition~\cite{smm,muelken01,lin19}, in finite systems while the percolation phase transition is considered to be of the continuous type.

\begin{figure}[htb!]
	\centering
	\includegraphics*[width=56mm]{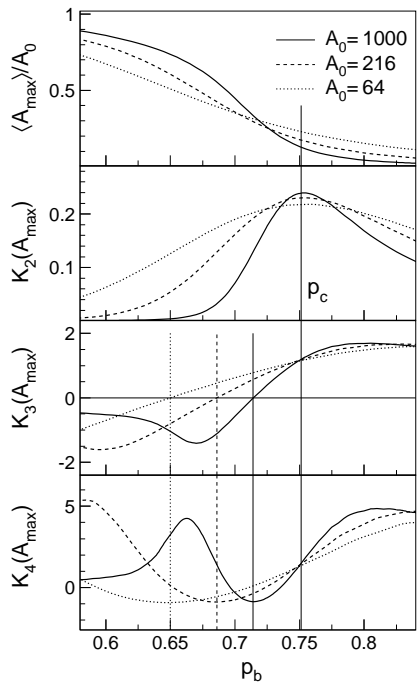}
	\vskip -0.1cm
	\caption{The normalized mean maximum fragment size $A_{\rm max}$ 
		and the cumulant ratios of Eqs. (1) as a function of the bond breaking
		probability $p_{b}$, as obtained with bond percolation for three
		different system sizes $A_0 = 64$, 216, and 1000.
		The long vertical line indicates the critical point $p_{c}$
		in the continuous limit. The short lines indicate
		the transition (pseudocritical) points for the finite
		systems (adapted from Ref.~\cite{jb18}). 
	}
	\label{fig_1}
\end{figure}

The predictions of the bond percolation model for the distribution properties of the largest
fragment in systems of various magnitudes are shown as a function of the bond-breaking probability $p_{b}$ in Fig.~\ref{fig_1}.
A Monte-Carlo procedure is used to randomly generate bonds
with probability $p = 1-p_b$ between neighboring sites arranged on the simple cubic lattice in the most compact configuration for a given system size. Clusters are identified with 
the Hoshen-Kopelman algorithm~\cite{h-k} and free boundary conditions are applied to account for
the presence of a surface in real systems.

The figure illustrates how the normalized magnitude of the largest fragment decreases with increasing bond breaking probability $p_b$. The transition becomes smoother with decreasing size of the considered system. In the limit of the infinite system a sharp transition occurs at the critical value $p_b = 0.751$~\cite{bauer86,sta}. 

The dimensionless cumulant ratios are derived from the central moments $\mu_i=\langle(A_{\rm max}-\langle A_{\rm max}\rangle)^i\rangle$ with $\langle A_{\rm max}\rangle$ denoting the mean value of the largest fragment $A_{\rm max}$. 
The cumulants $\kappa_i$ are simple functions of the central moments of which
$\kappa_1 = \langle A_{max}\rangle$, $\kappa_2 = \mu_2$, $\kappa_3 = \mu_3$,
and $\kappa_4 = \mu_4 - 3\mu_2^2$ contain the most significant information about the distribution.
The cumulant ratios $K_i$ are defined as
\begin{eqnarray}
K_2\equiv &\mu_{2}/\langle A_{\rm max}\rangle^2& =\kappa_2/\kappa_1^2 \nonumber\\
K_3\equiv &\mu_{3}/\mu_{2}^{3/2}& =\kappa_3/\kappa_2^{3/2} \nonumber\\
K_4\equiv &\mu_{4}/\mu_{2}^{2}-3& =\kappa_4/\kappa_2^{2},
\end{eqnarray}
\noindent where
$K_{2}$ is the variance normalized to the squared mean, 
$K_{3}$ is the skewness indicating the distribution asymmetry, and
$K_{4}$ is the kurtosis excess measuring the degree of peakedness relative to the normal distribution for which $K_{4} = 0$.

The coincidence of the zero transition 
of the skewness and the minimum of the kurtosis excess with a value near -1 is very precise and persists if the size of the system is varied. The maximum of the normalized variance $K_2$ is located at or very close to the critical bond-breaking parameter~\cite{jb06,jb18}.

\begin{figure}[htb!]
\centering
\includegraphics*[width=100mm]{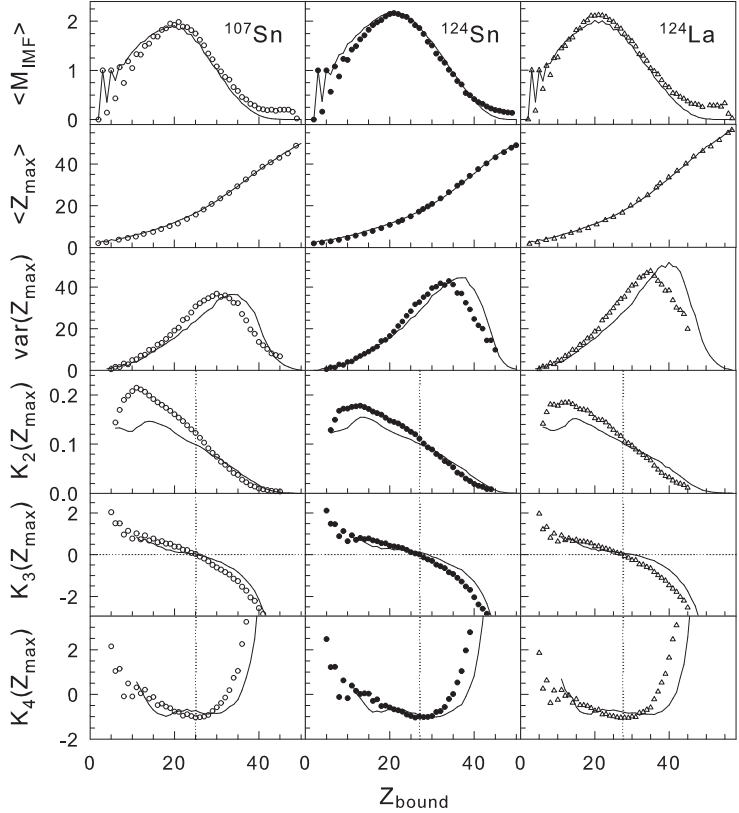}
\vskip -0.1cm
\caption{Experimental results (symbols) and SMM ensemble calculations (lines) for the mean multiplicity $\langle M_{\rm IMF} \rangle$ of 
intermediate-mass fragments, the mean value $\langle Z_{\rm max} \rangle$ of the largest fragment charge (both from Ref.~\protect\cite{ogul11}),
the variance $var (Z_{\rm max})$, and the results for the cumulant ratios $K_2$ to $K_4$, from top to bottom,
as a function of $Z_{\rm bound}$ for the three studied reactions $^{107,124}$Sn+Sn and $^{124}$La+Sn 
at 600 MeV/nucleon (from left to right). The results for $K_2$ are smoothed to suppress odd-even effects at small $Z_{\rm bound}$. The dotted vertical lines in the lower panels indicate the observed zero transitions of $K_3$.
}
\label{fig_2}
\end{figure}


The ALADIN experiment S254 was conducted at the GSI Helmholtzzentrum in Darmstadt. Beams of $^{107}$Sn, $^{124}$Sn and $^{124}$La 
were used to investigate isotopic effects
in projectile fragmentation at 600 MeV/nucleon. The radioactive secondary beams
with neutron-poor $^{107}$Sn and $^{124}$La projectiles contained also some fraction of neighboring isotopes. 
The mean compositions of the nominal $^{107}$Sn and $^{124}$La beams were 
$\langle Z \rangle$ = 49.7 and 56.8 and $\langle A/Z \rangle$ = 2.16 and 2.19, respectively~\cite{lukasik08}. 
Details of the experimental setup and the analysis are given in Ref.~\cite{ogul11}. The high efficiency of close to 100\% for
the detection of projectile fragments is an important property of the setup~\cite{sch96}. Tests performed for the case of
$^{197}$Au fragmentation confirmed that possible minor inefficiencies have negligible effects on the studied
cumulant ratios~\cite{jb18}. 

The most prominent result of the experiment is the observation that the isotopic dependence of projectile fragmentation is 
weak~\cite{sfienti09}. The results reported in Ref.~\cite{ogul11} for the mean multiplicity of intermediate-mass fragments (IMF, $3 \le Z \le 20$) and the mean atomic number $\langle Z_{\rm max} \rangle $ of the largest fragment in an event are shown in the top two rows of Fig.~\ref{fig_2}. The rise and fall of the IMF multiplicity as a function of $Z_{\rm bound}$ with maxima slightly exceeding the value $\langle M_{\rm IMF} \rangle = 2$ is very similar for the three cases. The same holds for the evolution of the mean value $\langle Z_{\rm max} \rangle $ and the variance $var (Z_{\rm max})$ displayed below. The quantity $Z_{\rm bound}$, defined as the sum of the atomic numbers $Z_i$ of all detected
fragments with $Z_i \geq$ 2 and chosen as the principal variable for event sorting, 
is monotonically correlated with the impact 
parameter of the reaction~\cite{sch96,hubele91}. 

The cumulant ratios shown in the lower three rows were extracted from the event-sorted data files used for the presentation of the experimental results in Ref.~\cite{ogul11}. It is evident that they exhibit the properties known from percolation~\cite{jb06}, qualitatively similar to
the case of the $^{197}$Au fragmentation discussed in Ref.~\cite{jb18}. 
In particular, the pseudo-critical points are rather precisely determined by the coincident zero transitions 
of $K_3$ and the minima of $K_4$ with values of about -1. For all three systems, they are located in the same small 
interval of $Z_{\rm bound}= 25$ for $^{107}$Sn to $Z_{\rm bound}= 27$ for $^{124}$Sn and $^{124}$La, 
approximately where $\langle M_{\rm IMF} \rangle$ rises most steeply with decreasing $Z_{\rm bound}$. 
The uncertainties caused by secondary deexcitations of the heaviest fragments were found to be small,
confirming $K_3$ and $K_4$ as robust indicators of the transition points~\cite{jb18}.


As reported by Ogul {\it et al.}~\cite{ogul11}, the interpretation of the data was performed within the Statistical Multifragmentation Model~\cite{smm}. An ensemble of hot sources representing the variety of excited spectator nuclei expected 
in a participant-spectator scenario was chosen with parameters determined empirically by searching for an optimum reproduction of the measured fragment charge distributions and correlations. 
The quality of the description achieved for the three studied reaction systems is illustrated in Fig.~\ref{fig_2}. The top two rows present the results reported in Figs. 13, 14 of Ref.~\cite{ogul11}. 

The cumulant ratios in the lower three rows show that the good quality of the reproduction extends to the higher-order properties of the $Z_{\rm max}$ distributions. Deviations are observed for $K_2$ in the region of smaller $Z_{\rm bound}$ and for $K_3$ and $K_4$ at larger $Z_{\rm bound}$. A precise agreement is observed near and around the pseudo-critical point that is equally well determined by the SMM results. As documented in Ref.~\cite{ogul11}, the evolution of the fragment $Z$ spectra with $Z_{\rm bound}$ exhibits the well-known transition 
from U-shaped through power-law to exponential spectral forms. The location of the identified transition points falls into the $Z_{\rm bound}$ interval with power-law shaped $Z$ spectra.

\begin{figure}[htb!]
\centering
\includegraphics*[width=80mm]{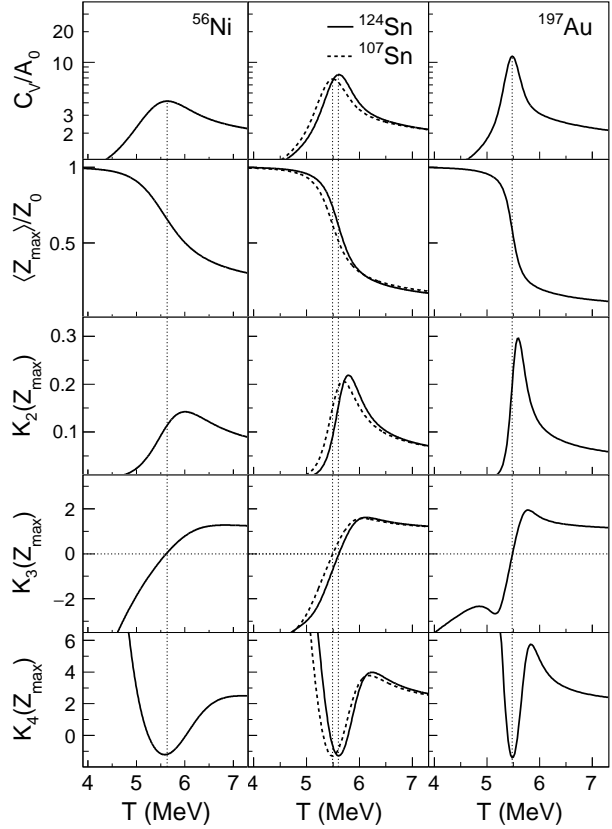}
\vskip -0.1cm
\caption{The normalized heat capacity $C_{V}/A_0$, the mean value $\langle Z_{\rm max} \rangle$ of the largest fragment charge normalized with respect to the system charge $Z_0$, and the results for the cumulant ratios $K_2$ to $K_4$ of the distributions of 
the largest fragment charge $Z_{\rm max}$, from top to bottom, as a function of the temperature, as calculated with the 
two-component thermodynamic model for the indicated nuclear systems. The vertical dotted lines indicate the zero transitions of $K_3$ seen to
coincide with the minima of $K_4$ and the maxima of the heat capacity.
}
\label{fig_3}
\end{figure}

The statistical errors are omitted in the figure but may be estimated from the scatter of the data points and lines. The significant odd-even effects visible at small $Z_{\rm bound}$ in the multiplicities of intermediate-mass fragments (top row of Fig.~\ref{fig_2}) are caused by the definition of $Z_{\rm bound}$ that includes He fragments. Small even values of $Z_{\rm bound}$ may thus contain larger amounts of multiple-$^{4}$He events without a fragment with $Z \ge 3$. This is reflected in the cumulant ratios $K_3$ and $K_4$ with slightly more peaked and more skewed distributions for the smaller odd values of $Z_{\rm bound}$. In the $K_2$ distributions, this effect has been suppressed by smoothing over the two neighboring values of $Z_{\rm bound}$ with weights [0.25, 0.5, 0.25]. It permits the maxima of $K_2$ at $Z_{\rm bound} = 11$ for $^{107}$Sn and $Z_{\rm bound} = 13$ for the two heavier projectiles to appear more clearly. These values may indicate approximate locations of the true critical point as discussed in Ref.~\cite{jb18}. 
The differences of $K_2$ at small $Z_{\rm bound}$, between experimental and model results and between systems, are not surprising because the variances and values of $\langle Z_{\rm max} \rangle$ are small and the tiny overprediction of $\langle Z_{\rm max} \rangle$ and underprediction of the variances in 
some cases have visible effects on $K_2$.


The appearance of a percolation-type pseudocritical transition in the $Z_{\rm max}$ distributions within the SMM model suggests this as a generic phenomenon in fragmentation processes. It may be present in other models as well which here is tested by presenting calculations with the canonical thermodynamic model (CTM) proposed by Das {\it et al.} and known to contain a first-order phase transition~\cite{das05}. The thermodynamic model is a simplified version of the statistical fragmentation model. A recursion relation permits computing the partition function and thus to obtain thermodynamic properties of the system. It may include the isospin degree of freedom and Coulomb forces as in the two-component version
that is used here. Calculations are performed for a system of given mass $A_0$, mass over charge ratio $A_0/Z_0$ and temperature $T$. The freeze-out density has to be specified and $\rho_0 /3$ with $\rho_0 = 0.16$~fm$^{-3}$ was used.

The results of CTM calculations with standard parameters~\cite{das05} for the four nuclear systems $^{56}$Ni, $^{107}$Sn, $^{124}$Sn, and $^{197}$Au are shown in Fig.~\ref{fig_3}. In addition to the cumulant ratios $K_2$ to $K_4$ and the mean normalized atomic number $\langle Z_{\rm max}\rangle / Z_0$, also the normalized heat capacity $C_V/A_0$, with $C_V = {\partial E} / {\partial T}$ at constant volume, is presented in the figure. Defining the transition temperature through the zero crossing of the skewness $K_3$, we observe that it coincides with the minimum of $K_4$, as expected from percolation, but also with the maximum of the heat capacity $C_V$. The variation of $\langle Z_{\rm max}\rangle$ is fastest at the transition point as known from Refs.~\cite{jb18,gupta98,das18,das05}. The effect of varying the neutron content of the Sn isotopes is small. 
It is evident that the thermodynamic model exhibits the same critical features that, in this model, coincide with the maximum of the specific heat
expected to be a trace of a phase transition in small systems. The corresponding temperature has been named boiling point~\cite{gupta98,das05}.

\begin{table}[ht]
\begin{center}
\begin{tabular}{|c|c|c|c|c|c|c|}
\hline
Projectile  & $Z_{\rm bound}$ & $Z_0$ & I & $T_{\rm HeLi}$ & $T_{\rm SMM}$ & $T_{\rm CTM}$ \\ 
&  &  &  & (MeV) & (MeV) & (MeV) \\
\hline
$^{107}$Sn & 25 & $30 \pm 2$ & 0.074 & 5.4 & 6.0 & 5.5 \\
$^{124}$Sn & 27 & $32 \pm 2$ & 0.194 & 5.8 & 5.9 & 5.6 \\
$^{124}$La & 27 & $32 \pm 2$ & 0.086 & 5.5 & 6.0 & 5.5 \\
\hline
\end{tabular}
\end{center}
\caption{Freeze-out characteristics of the three fragmenting systems at the transition points $Z_{\rm bound}$: the estimated atomic number $Z_0$, the asymmetry $I=(N-Z)/A$ of the projectile (averaged over the beam composition for the radioactive projectiles), the measured apparent temperature $T_{\rm HeLi}$ at the transition point from Ref.~\protect\cite{sfienti09}, the mean microcanonical temperature $T_{\rm SMM}$ from Ref.~\protect\cite{ogul11}, and the CTM freeze-out temperature calculated for the indicated system charge $Z_0$ and asymmetry $I$. 
}
\label{tab_1}
\end{table}

The presented calculations were performed for the given nuclei and varying temperatures (Fig.~\ref{fig_3}). With a parametrization of the impact parameter dependent magnitudes of the projectile spectator mass, charge, and temperature, Mallik {\it et al.} have achieved a realistic description of the main observables of the present reactions with CTM calculations~\cite{mallik11}. It will be interesting to know whether also there the fluctuations of the largest fragment charge exhibit the percolation features. The agreement of so defined transition points with the experimental result shown in Fig.~\ref{fig_2} would provide additional support for the theoretical approach.


Overall, the dependence of the studied observables on the isotopic composition of the initial projectile system is very small. This holds for the experimental as well as for the theoretical results calculated with the SMM and CTM (Figs.~\ref{fig_2},\ref{fig_3}). 
The investigated higher-order fluctuations of the largest fragment charge thus confirm the earlier observations (Refs.~\cite{sfienti09,ogul11,buyuk05}) that the fragmentation process is governed by the opening of the corresponding partition space and less likely by a possible Coulomb instability related to the magnitude of the charge residing in the system.

\begin{figure}[htb!]
\centering
\includegraphics*[width=80mm]{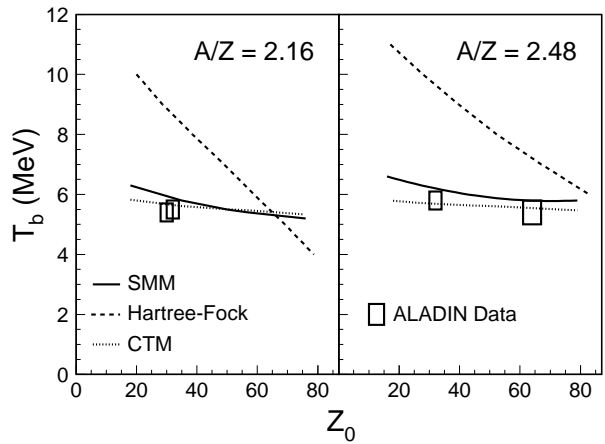}
\vskip -0.1cm
\caption{Breakup temperatures $T_b$ as a function of the reconstructed atomic number $Z_0$ of the disintegrating system
at the pseudocritical breakup point for neutron-poor (left) and neutron-rich 
(right) systems. The boxes represent the apparent chemical breakup temperatures $T_{\rm HeLi}$ from Table~\protect\ref{tab_1} with an assumed error of $\pm 0.3$~MeV while the lines give the model results as indicated. The value $5.4 \pm 0.4$~MeV for the fragmentation of $^{197}$Au~\protect\cite{temp}  is included in the right panel.  
}
\label{fig_4}
\end{figure}

Starting from the location of the transition points on the $Z_{\rm bound}$ axis (Fig.~\ref{fig_2}), the parameters of the fragmenting sources were evaluated (Table~\ref{tab_1}). The mean atomic number
$Z_0$ was estimated on the basis of percolation as being larger by $\approx 20\%$ than the observed $Z_{\rm bound}$ value at the transition (cf. Fig. 4 in Ref.~\cite{jb18}). The isotopic asymmetry $I=(N-Z)/A$ is assumed to be that of the initial projectiles. The values of the isotope temperature $T_{\rm HeLi}$ were taken from Ref.~\cite{sfienti09}, and the so-called apparent temperatures at the $Z_{\rm bound}$ value of the transition points were used. 
Model calculations indicate that the apparent $T_{\rm HeLi}$ may underestimate the actual breakup temperature by up to 15\% because of the effects of secondary decays~\cite{temp} and corrections of 20\% were used in Refs.~\cite{sfienti09,natowitz02}. 
Here the observed values are found to be on average 0.4 MeV, 
i.e. less than 10\%, 
lower than the mean microcanonical temperatures obtained with the SMM for the same breakup points~\cite{ogul11}. The statistical errors of $T_{\rm HeLi}$ are below 0.1 MeV~\cite{sfienti09}, so that the experimental results may suggest a minor dependence on the isotopic composition of the system. The measured values are lower by $\approx 0.4$~MeV in the case of the neutron-poor, i.e. more highly charged, systems. This is not reflected in the calculated SMM and CTM results (Table~\ref{tab_1}) nor is it evident from the apparent $T_{\rm BeLi}$ isotope temperatures of 5.6 to 5.7~MeV reported in Ref.~\cite{sfienti09}. 
The CTM temperatures for the specified systems at their boiling points are similar to those of the nominal projectiles displayed in Fig.~\ref{fig_3} because their mass dependence is very small.  

The isotopic and system size dependence of the breakup temperature is shown again in Fig.~\ref{fig_4}, together with the expectations obtained with the SMM and CTM as well as from the Hartree-Fock calculations reported by Besprosvany and Levit~\cite{besprosvany89}. For this purpose, SMM and CTM calculations were performed for many systems covering the interval of $ 20 \lesssim Z_0 \lesssim 80$, for the two indicated asymmetries $I$, and by generating events over wide ranges of either the excitation energy (SMM) or the temperature (CTM). The transition points were identified on the basis of the obtained cumulant ratios by searching for their characteristic signal. In the case of the SMM, the microcanonical breakup temperatures derived in this way for the experimentally studied systems are within $\approx 0.2$~MeV the same as the results of the ensemble calculations listed in the table.

\begin{figure}[htb!]
\centering
\includegraphics*[width=75mm]{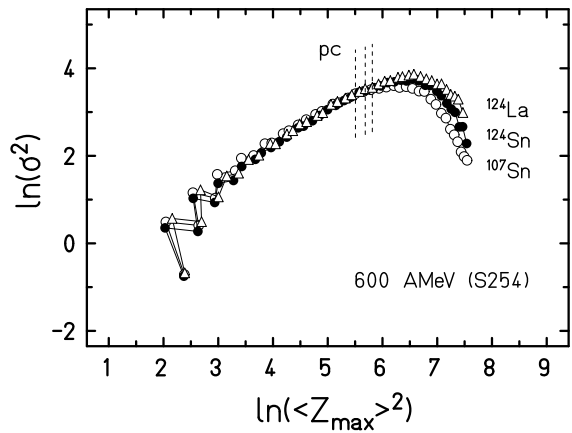}
\vskip -0.1cm
\caption{Natural logarithm of the variance as a function of the natural logarithm of the squared mean value of the largest atomic number
$Z_{\rm max}$ recorded in the three reactions with $^{107}$Sn (open circles), $^{124}$Sn (filled circles), and $^{124}$La (open triangles)
projectiles at $600$ MeV/nucleon. The data symbols represent the results for individual values of $Z_{\rm bound}$ in the range
from 5 to 45. The positions of the pseudocritical (pc) points are indicated by dashed vertical lines and correspond, from left to right, to the
$^{107}$Sn, $^{124}$Sn, and $^{124}$La systems. 
}
\label{fig_5}
\end{figure}

The results for the neutron-poor and the neutron-rich cases are shown separately in the two panels. The experimental 
temperatures $T_{\rm HeLi}$ are given with an assumed error of $\pm 0.3$~MeV and include $T_{\rm HeLi} = 5.4 \pm 0.4$~MeV for $^{197}$Au at $Z_{\rm bound} = 54$ from Ref.~\cite{temp}, corresponding to $Z_0 = 64$ (Ref.~\cite{jb18}). After applying the same 20\% correction, these values are in very good agreement with the plateau temperatures of caloric curves compiled by Natowitz {\it et al.}~\cite{natowitz02}. 
It is evident that the fragmentation occurs at temperatures lower than those expected for the Coulomb instabilities of equilibrated compound nuclei investigated with the Hartree-Fock model~\cite{besprosvany89}. The magnitudes of the latter may still depend on the form of the nuclear potential used in the calculations~\cite{levit85,baldo04} but the predicted dependences on the mass and isotopic composition of the system are much stronger than in the statistical model cases and as observed here and in other fragmentation reactions~\cite{liu19}.


The fluctuations of the largest fragment size observed for $^{107,124}$Sn and $^{124}$La fragmentations at 600 MeV/nucleon support the earlier results derived from the 
analysis of $^{197}$Au fragmentations at similar energies and confirm the conclusions derived there~\cite{jb18}. 
For example, it was shown that no evidence for $\Delta$ scaling~\cite{botet01,frankland05} was found in the $^{197}$Au spectator fragmentation.
As one of its consequences, a linear correlation of the natural logarithms of the variance and of the squared mean value of the largest atomic number $Z_{\rm max}$ would be expected but is not observed. The variation of the slopes is continuous and very smooth also in the present case (Fig.~\ref{fig_5}). The correlations recorded for the three reactions are nearly identical and also very similar to that reported for $^{131}$Xe + $^{27}$Al at the same energy in Ref.~\cite{jb18}. This includes the odd-even effects at small ln($\langle Z_{\rm max}\rangle^2)$ known to be caused by the definition of $Z_{\rm bound}$ (cf. Fig.~\ref{fig_2}). Qualitatively similar shapes have been reported for percolation~\cite{jb06} and for the canonical lattice gas model with constant density and varying temperature~\cite{gulm2005}.

As a main result, it is found that the SMM model in the form used to successfully describe the fragment distributions and correlations of the studied reactions reproduces also the fluctuations of the largest fragment size up to fourth order. An observation of comparable kind has been made in an earlier analysis of the fragmentation of $^{197}$Au on Cu targets at 600 MeV/nucleon~\cite{botvina95}. Distributions of $Z_{\rm max}$ for finite intervals of $Z_{\rm bound}$ or of the IMF multiplicity were found to be quantitatively reproduced, with structures beyond their widths given in detail, after a parametrization of the model parameters based on global observables had been achieved. It points to the role of internal constraints being present in fragmentation processes.

Here the evolutions of the experimental and calculated cumulant ratios with $Z_{\rm bound}$ were seen to track each other rather well, and even very precisely at and near the transition points. 
This fact is remarkable because the
pseudocritical points indicated by $K_3=0$ and a minimum of $K_4$ are known to be a property of the continuous phase transition in percolation~\cite{jb06}. 
The SMM description, on the other hand, emulates many properties of the nuclear reaction. The variation with impact parameter is modeled by assuming an ensemble of hot sources and, after the initial partitioning, the deexcitation of excited fragments is followed realistically with a variety of descriptions adapted to the fragment mass and excitation energy. None of these elements of the reaction scenario are present in percolation. The only common features of the two models are the short range of the interaction of the constituents, the three-dimensional space in which the systems evolve, and the stochastic nature of the elementary processes permitting Monte-Carlo methods for numerical realizations. 

The same properties of the higher-order cumulant ratios were seen to be present also in the CTM, with the percolation-type transition found at the boiling point of the phase transition described with that model~\cite{das05}.
One may conclude that they should be generic for disintegration processes. 
Related suggestions were, e.g., made by Biro {\it et al.}~\cite{biro86} and by Campi {\it et al.}~\cite{campi00} who, 
in their search for a universal mechanism of fragmentations in simple fluids, have argued that the random breaking of bonds 
may be the simplest explanation for the appearance of percolation features in nuclear fragmentation.

With this conclusion, one may expect that they are present as well in dynamical descriptions of the same type of reaction. Studies based on quantum-molecular dynamics (QMD) transport models have demonstrated that the asymptotic cluster partition can be recognized at early reaction times, right after the violent initial collision rate has subsided and equilibrium is still in the process of being reached~\cite{gossiaux97,lefevre19}. The present $^{107,124}$Sn and $^{124}$La fragmentations have been successfully reproduced with the isospin-dependent quantum molecular dynamics (IQMD) model of Su {\it et al.}~\cite{junsu18} who show reaction trajectories in the temperature vs density plane in a subsequent paper~\cite{junsu19}. If the expansion to subsaturation densities reported there proceeds approximately adiabatic~\cite{bertsch83}, one will have to conclude that the early partitioning, primarily in momentum space, is already occurring during the high-density phase of the reaction. The persistence of the fragmentation patterns during expansion has been emphasized in QMD~\cite{gossiaux97,lefevre19} and also in classical molecular dynamics studies~\cite{campi03}. 
It will be interesting to see the higher-order cumulant ratios that are obtained with these dynamical calculations.


In summary, the experimental data for the fragmentation of Sn and La projectiles at 600 MeV/nucleon and calculations with the SMM and CTM statistical models corroborate the suggestion derived from percolation that the cumulant ratios skewness $K_3$ and kurtosis excess $K_4$ of the distributions of the largest fragment charge are valuable observables in searching for a phase transition in multifragmentation. The location of the transition is indicated by $K_3 = 0$ coinciding with a minimum of $K_4$. The SMM and CTM calculations reproduce this transition very well that, according to the  CTM, coincides with the maximum of the heat capacity $C_V$.  Signatures conventionally associated with first- and second-order transitions in the continuous limit apparently coexist in finite systems. The observed properties of the higher-order cumulant ratios appear generic for fragmentation processes. It will be interesting to explore their existence in other fragmentation models including dynamical descriptions of the process.

The authors would like to express their gratitude to the ALADIN Collaboration for the
permission to use the event-sorted fragmentation data of experiment S254. Two of us (A. S. B. and N. B.) are grateful for the hospitality received at the Frankfurt Institute for Advanced Studies (FIAS), Frankfurt am Main, Germany. N. B. acknowledges support by the Scientific and Technological Research Council of Turkey (TUBITAK) under Project No. 118F111 and within the
framework of COST Action CA15213 THOR. 
This work was supported by the Polish Scientific Research Committee, Grant No. 2P03B11023, by the European Community under Contract No. HPRI-CT-1999-00001, and by the Polish Ministry of Science and Higher Education Grant N202 160 32/4308 (2007-2009). 

\end{document}